\documentstyle[12pt]{article}
\begin{document}
\hsize = 6.0 in
\vsize =11.7 in
\hoffset=0.1 in
\voffset=-0.5 in
\baselineskip=20pt
\newcommand{\ghat}{{\hat{g}}}
\newcommand{\Rhat}{{\hat{R}}}
\newcommand{\ih}{{i\over\hbar}}
\newcommand{\Scal}{{\cal S}}
\newcommand{\fudge}{{1\over16\pi G}}
\newcommand{\tn}{\mbox{${\tilde n}$}}
\newcommand{\mg}{\mbox{${m_g}^2$}}
\newcommand{\mf}{\mbox{${m_f}^2$}}
\newcommand{\hk}{\mbox{${\hat K}$}}
\newcommand{\vk}{\mbox{${\vec k}^2$}}
\newcommand{\eqletter}{ \hfill (\theequation\alph{letter})}
\newcommand{\gm}{{(\Box+e^2\rho^2)}}
\newcommand{\eql}{\nonumber &\eqletter \cr
                  \addtocounter{letter}{1}}
\newcommand{\be}{\begin{equation}}
\newcommand{\ee}{\end{equation}}
\newcommand{\bea}{\begin{eqnarray}}
\newcommand{\eea}{\end{eqnarray}}
\newcommand{\beal}{\setcounter{letter}{1} \begin{eqnarray}}
\newcommand{\eeal}{\addtocounter{equation}{1} \end{eqnarray}}
\newcommand{\none}{\nonumber \\}
\newcommand{\req}[1]{Eq.(\ref{#1})}
\newcommand{\reqs}[1]{Eqs.(\ref{#1})}
\newcommand{\larrow}{\,\,\,\,\hbox to 30pt{\rightarrowfill}
\,\,\,\,}
\newcommand{\slarrow}{\,\,\,\hbox to 20pt{\rightarrowfill}
\,\,\,}
\newcommand{\half}{{1\over2}}
\newcommand{\bfx}{{\vec{x}}}
\newcommand{\bfy}{{\vec{y}}}
\newcommand{\zfp}{Z_{{FP}}}
\newcommand{\zf}{Z_{{F}}}
\newcommand{\zr}{Z_{{R}}}
\newcommand{\zop}{Z_{{OP}}}
\newcommand{\zekt}{Z_{EKT}}
\newcommand{\phstar}{{\varphi^\dagger}}
\begin{center}
{\bf Quantum Theory of Black Holes}\\
\vspace{15 pt}
{\it by}\\
\vspace{13 pt}
J. Gegenberg${}^1$ {\it and}  G. Kunstatter${}^2$\\[5pt]
${}^1${\it Department of Mathematics and Statistics}\\
   {\it University of New Brunswick}\\
   {\it Fredericton, New Brunswick}\\
   {\it CANADA E3B 5A3}\\[5pt]
  ${}^2${\it  Institute for Theoretical Physics and Physics
Department}\\
{\it University of Winnipeg}\\
{\it Winnipeg, Manitoba}\\
{\it CANADA, R3B 2E9}\\
\end{center}
\vspace{40pt}
{\narrower\smallskip\noindent
{\bf Abstract} :
A solvable 2-dimensional conformally invariant midi-superspace
model for black holes is obtained by imposing spherical symmetry
in 4-dimensional conformally invariant Einstein gravity.  The
Wheeler-DeWitt equation for the theory is solved exactly to
obtain the  unique quantum wave functional for an isolated black
hole with fixed mass. By suitably relaxing the boundary
conditions, a non-perturbative ansatz is obtained for the wave
functional of a black hole interacting with its surroundings.}
\vspace{40 pt}
\begin{center}
{\it November, 1992}
\end{center}
\par
\vspace*{20pt}
\noindent
WIN-92-9
\noindent
UNB Technical Report 92-05
\clearpage
One of the most important developments in field theory in the
last two decades was the discovery of  the quantum
mechanical instability of black holes due to Hawking
radiation\cite{Hawking}. This discovery
provided a tantalizing, and still poorly understood, link between
two previously distinct branches of physics, namely gravitation
theory and thermodynamics\cite{Beckenstein}. In addition,
questions surrounding
the endpoint of black hole radiation have touched on the
foundations of our understanding of both quantum mechanics and
thermodynamics\cite{Preskill}.
\par
Most calculations of black hole radiation involve matter fields
quantized on a
classical curved background. Recently the  backreaction of the
quantum matter fields on the gravitational field has been studied
semi-classically
in a class of 2-D models which exhibit many features in common
with
4-D gravity\cite{CGHS}. Another interesting method\cite{york} for
studying black hole
thermodynamics uses the Euclidean action
for black holes to approximate functional integral expressions
for the relevant thermodynamic partition functions.
Unfortunately, although many interesting results have been
obtained, neither approach
has yet provided a resolution to the question of the endpoint of
gravitational
collapse, which appears to lie outside the realm of validity of
the semi-classical approximation.
\par
The purpose of this Letter is to present a completely different
and inherently non-perturbative approach to these issues. In
particular, we
study a two dimensional conformally invariant midi-superspace
model for
black holes in which the gravitational field can  be quantized
exactly. The analysis of the exact quantum
theory for such a model can in
principle provide information about the validity of the
semi-classical
approximation, the significance of backreaction effects, and
ultimately the
nature of the endpoint of decay by Hawking radiation.
The model is obtained by
imposing spherical symmetry in conformally invariant
4-dimensional Einstein
gravity\cite{GK1}. It is important to stress that the 4-d theory
is classically equivalent to Einstein gravity so that the model
in principle makes direct contact with physical, four dimensional
black holes. A semi-classical analysis has shown that the
``matter fields" in the theory give rise to Hawking radiation
with
the usual temperature\cite{GK1}. A related model has
also
been analyzed\cite{Mohammedi,Lowe} using the methods of
Ref.\cite{CGHS}.
In what follows, the theory will be quantized exactly using
techniques
first applied by Henneaux\cite{Henneaux} to
Jackiw-Teitelboim 2-D gravity\cite{twodim}.
\par
We start from the classical action for a scalar field conformally
coupled to gravity in four dimensions:
\be
I^{(4)}[\phi,\ghat_{ab}]= \kappa\int d^4{\hat x} \sqrt{-\ghat}
\left(\phi^2\Rhat+6\ghat^{ab}{\hat \nabla}_a\phi {\hat
\nabla}_b\phi \right),
\label{eq: 4d action}
\ee
where $\{a,b=0,1,2,3\}$ and $\kappa=\fudge$. Without loss of
generality, we will normalize the vacuum expectation value of
the scalar field to unity.
\req{eq: 4d action} is invariant under the conformal
transformations: $\ghat_{ab}\to e^{2\sigma}\ghat_{ab}$ and
$\phi \to e^{-\sigma}\phi$. The theory
is equivalent to Einstein gravity classically, and
(perturbatively) at the  quantum level as well\cite{Tsamis}.
\par
A midi-superspace model for black holes is obtained by imposing
exact spherical symmetry with 4-metric:
\be
ds^2=g_{\mu\nu}(x)dx^\mu dx^\nu + \lambda^2(x) d\Omega^2,
\ee
where the fields, including the matter field $\phi$, are now
functions only of $x^\mu = \{r,t\}$ and
$d\Omega^2$ is the standard line element on the two sphere with
volume $4\pi$.
The reduced action
\be
I^{(2)}= 24\pi\kappa \int d^2x \sqrt{-g}\left({1\over3}\tau R(g)+
g^{\mu\nu}\nabla_\mu\tau \nabla_\nu\psi +
{1\over3} {e^{3\psi}\over\sqrt{2\tau}}\right),
\label{eq: 2d action}
\ee
describes a two dimensional, conformally invariant field theory
with two ``matter" fields $\tau:=\half\lambda^2\phi^2$
and $e^{3\psi}:=\lambda\phi^3$\cite{dimensions}.
In terms of this parametrization, conformal transformations take
$g_{\mu\nu}\to e^{2\sigma} g_{\mu\nu}$ and $\psi\to
\psi - {2\over3}\sigma$, while $\tau$ is
invariant. The field equations obtained by varying \req{eq: 2d
action} are equivalent to those obtained by imposing spherical
symmetry on the equations obtained from the four dimensional
action (\ref{eq: 4d action})\cite{KaluzaKlein}.  Consequently,
Birkhoff's
theorem in
4-dimensions guarantees that the 2-dimensional theory is
classically
solvable. Up to diffeomorphisms and conformal
transformations, there exists only a one parameter family of
solutions:
\bea
\tau&=&\half r^2,\none
e^{3\psi}&=& r,\none
ds^2&=& -(1-2m/r)dt^2 + (1-2m/r)^{-1}dr^2.
\label{eq: 2d solution}
\eea
These solutions describe black holes of mass $m$, with 4-
dimensional dilaton $\phi^2=1$.
\par
The theory based on the action (\ref{eq: 2d action}) was first
used in \cite{GK1} to derive an
expression for Hawking radiation in a semi-classical
approximation by computing the trace anomaly of the ``matter"
fields $\tau$ and $\psi$. Here we present the results of the
exact
quantization of the full theory, using the methods of
Henneaux\cite{Henneaux}. Details will be given
elsewhere\cite{GK2}.
\par
 As with all diffeomorphism invariant theories\cite{HRT},
the Hamiltonian is, up to a surface term, a linear combination of
constraints:
\be
H=\int dr \left\{{1\over 2G}\sigma {\cal G} +M {\cal F}
+\Lambda\Pi_\beta
\right\}+H_{ADM}
,\ee
where the primes denote differentiation with respect to $r$,
and we have
defined the fields
$\alpha:=2\rho+3\psi$ and $\beta:=2\rho-3\psi $ where
$e^{2\rho}=g_{11}$ represents the conformal mode of
the 2-metric in our parametrization. The field $\alpha$ is
conformally
invariant while the ``pure
(conformal) gauge" component $\beta$ has disappeared from the
Hamiltonian, as
required. $\Pi_\alpha$, $\Pi_\tau$ and $\Pi_\beta$ are momenta
canonically conjugate to $\alpha$, $\tau$ and $\beta$
respectively. In
this parametrization the generator of conformal transformations
is simply $\Pi_\beta$ and the conformal mode can be trivially
eliminated without affecting the subsequent discussion.
\par
The Lagrange multipliers $\sigma$ and $M$ are related to
the
lapse and shift functions, and  the constraints:
\bea
{\cal F}&=&   \alpha'\Pi_\alpha+\tau'\Pi_\tau
-2\Pi'_\alpha\approx 0,\\
{\cal G}&=&  2\tau'' -\alpha'\tau' - {G^2\over
4}\Pi_\alpha\Pi_\tau
   -{e^\alpha\over\sqrt{2\tau}}\approx 0,
\label{eq: constraints}
\eea
generate spatial diffeomorphisms, and time translations,
respectively. The ADM energy\cite{HRT} is:
\be
H_{ADM}= {1\over 2G}\int dr (\sigma \tau'- 2\sigma\alpha'\tau)'.
\label{eq: adm energy}
\ee
It can easily be verified that for the solutions given in
\req{eq: 2d
solution}, $H_{ADM}=m/G$\cite{quantum hair}. Moreover, the
generator of time
translations given above is well defined for all configurations
which approach (\ref{eq: 2d solution}) asymptotically.
\par
As in Ref.\cite{Henneaux} we avoid potential factor
ordering problems associated with quadratic momentum contraints
by first solving the constraints classically. The result is:
\bea
\Pi_\alpha &=&  {1\over 2G} Q(\alpha,\tau),    \\
\Pi_\tau &=&   {1\over 2G} \displaystyle{{ (2\tau''-\alpha'\tau'-
e^\alpha/\sqrt{2\tau})\over Q(\alpha,\tau)}},
\label{eq: classical momenta}
\eea
where
\be
Q:= \left( (\tau')^2 +(C-\sqrt{2\tau})e^\alpha\right)^\half.
\label{eq: Q}
\ee
The
parameter $C$ is a constant of integration that determines the
allowed classical, static solutions: they are of the form
\req{eq: 2d solution} with $m=C/2$.
\par
This completes the discussion of the essential classical features
of the model. The quantum theory in the
functional Schrodinger representation will now be constructed
using the so-called
``Dirac approach", in which one first quantizes
the theory in the unreduced configuration space, and then imposes
the constraints as operator constraints on physical states. The
states in the unreduced theory are arbitrary functionals
$\psi[\alpha,\tau]$ of the
fields $\alpha(r)$ and
$\tau(r)$.
Conjugate momenta are defined
as
functional derivatives:
\bea
\hat{\Pi}_\alpha&:=& -i\hbar
\displaystyle{{\delta\over\delta\alpha(r)}}\nonumber\\
\hat{\Pi}_\tau&=& -i\hbar
\displaystyle{{\delta\over\delta\tau(r)}}
\label{eq: momentum operators}
\eea
These operators are formally self-adjoint with respect to the
inner product\cite{technical stuff}:
\be
<\psi|\psi>:=\int \prod_r [d\alpha(r)][d\tau(r)]
\psi^*[\alpha,\tau]\psi[\alpha,\tau]
\label{eq: measure}
\ee
\par
In the Dirac approach
physical states are functionals $\psi[\alpha,\tau]$ that
obey the operator constraints, which now take the form:
\bea
-i\hbar {\delta\over\delta\alpha(r)}\psi[\alpha,\tau]
&=& {1\over 2G} Q(\alpha, \tau) \psi[\alpha,\tau]\none
-i\hbar {\delta\over\delta\tau(r)}\psi[\alpha,\tau],
&=& {1\over 2G} \displaystyle{{ (2\tau''-\alpha'\tau'-
e^\alpha/\sqrt{2\tau})\over Q(\alpha,\tau)}}\psi[\alpha,\tau].
\eea
These equations can be functionally integrated (see
ref.\cite{GK2}) to yield the
unique (up to total divergences) physical state in the
theory:
\be
\Psi[\alpha,\tau]= \exp{i\over m_{pl}^2}\int dr \left\{ Q
+ {\tau'\over2}
   \ln\left({\tau'-Q\over\tau'+Q}\right)\right\}
,\label{eq: wave function}
\ee
where $m_{pl}=\sqrt{\hbar G}$ is the Planck length.
\par
This solution, which is one of the main results of this paper,
has several interesting properties: It is invariant
under spatial diffeomorphisms, and (trivially) under conformal
transformations. In addition, $\psi =1$ for the classical
solution in \req{eq: 2d solution}.  Moreover,
classically forbidden field configurations which have imaginary
momenta ($Q^2<0$: cf \req{eq: classical momenta}) yield wave
functions whose amplitudes are exponentially damped. Finally, we
note that  if the fields $\alpha(r)$ and $\tau(r)$ obey suitable
boundary conditions as $r\to\infty$, namely  $\tau(r)\to \half
r^2 (1+
O(1/r^2))$ and $\alpha(r)\to r+2m +O(1/r)$, then the state
\req{eq: wave function} is an
eigenstate of the ADM hamiltonian with eigenvalue $m/G=C/2G$.
These boundary conditions
effectively restrict consideration to fields  configurations
with classical ADM energy equal to $m/G$.
 \par
As usual in quantum gravity, physical interpretation of the
wavefunction \req{eq: wave function} requires considerable care.
For one thing, since the inner product on the Hilbert space is
given by a functional integral it is not obvious that the state
is normalizable (even after all field configurations related by
spatial diffeomorphisms are factored out). However, since it is
the {\bf only} state in the physical Hilbert space, we will
assume that its associated probability amplitude does in
principle contain information about relative probabilities of
quantum mechanically allowed field configurations.
\par
Another important point is the fact that the quantum theory as
constructed contains no physical degrees of freedom, and hence no
unconstrained observables. The solution as given above therefore
cannot directly yield information about Hawking radiation, or
gravitational collapse. In order to gain insight into these
questions, it is necessary to know how matter can ultimately be
incorporated into the model. In the following we will present an
ansatz that attempts to mimic the effect of interactions with its
surroundings by putting the black hole in  a box of radius
$R>>m$, and relaxing the boundary conditions on the fields to
include configurations that have ADM energy $M\neq m$. For
simplicity
we neglect local fluctuations and restrict consideration to
fields of the form:
\bea \tau= \half r^2,\none
e^\alpha=\displaystyle{{r^2\over(r-2M)}}, \eea
with support only
in
the region $r>r_0$, where $r_0<<2m$.  These fields correspond to
black holes of mass $M$, expressed in Schwarzschild coordinates.
For
fixed $C=2m$, they do not correspond to classical solutions
unless
$M=m$.  We will now evaluate the wave functional \req{eq: wave
function} for these configurations. The quantum state is
therefore a function of the single
variable
$M$. As one might anticipate, by relaxing the boundary conditions
so as to
allow exchange of energy with an external source, we have
abandoned the self-adjointness of the Hamiltonian: its action
takes states out of the physical Hilbert space. Nonetheless we
will see that the resulting wave function has some interesting
properties.
\par
The physically relevant
information
in the wave function\cite{mess} is contained in the
(unnormalized)
probability
amplitude $P[M]:= \exp ( -{2\over \hbar} Im S[M])$.  For $M>m$
\be
P[M]=\exp\left(-{3\over4 m_{pl}^2} (2M-2m)^{3\over2}\sqrt{R} +
O(m/R)
\right) ,
\label{eq: pm1}
\ee
and for $0\leq M<m$:  \bea P[M]&=&
\exp\left(-4{m^2\over
m_{pl}^2} \left[{\pi\over2} - \arctan{\sqrt{
 M\over m-M}}+\right.\right.\none &\,&\,\,\left.\left.- {M^2\over
m^2} \sqrt{m-M\over M} +\half
   {M (2m-2M)\over m^2}\sqrt{m-M\over M} \right]\right)
{}.
\label{eq: pm2}
\eea
Figure 1 contains graphs of the probability amplitude $P[M]$ for
different values of the classical mass $m$.  The simplifications
made
above yield a probability amplitude with remarkable properties:
\begin{itemize} \item The amplitude is finite and well behaved in
the
limit that $r_0\to 0$. The classical curvature singularity has
disappeared from the quantum amplitude.  \item The amplitude is
continuous and smooth (with zero slope) at $m=M$, as long as
$R\neq\infty$.  \item The amplitude is peaked at the classical
mass,
and the width decreases rapidly with increasing $m>m_{pl}$.
\item
The relative probability of configurations with mass $M>m$
(compared
to $M=m$) is exponentially supressed, with exponent proportional
to
the spatial volume.  \item The relative probability of
configurations
with mass $M=m$ and $M=0$ is \be {P[m]\over P[0]}= \exp{{2\pi}
{m^2\over m_{pl}^2}}
.\ee
\end{itemize} \par It is interesting to note that if one
interprets
this relative probability thermodynamically in terms of the
number of
(equally probable) microstates with mass $m$ (assuming a unique
zero
mass state), then one gets an expression for entropy for a black
hole
of mass $m$:  $S=k\ln{P[m]\over P[0]}= 2\pi m^2/m_{pl}^2$. This
is a
factor of two smaller than the standard value for the entropy of
a
black hole.
\par
Although Eqs.(\ref{eq: pm1},\ref{eq: pm2}) and their
interpretation are highly
speculative, the results outlined above seem to suggest that 2-D
models may provide a non-perturbative basis for the study of
black
hole radiation. In order to answer questions concerning
the
endpoint of gravitational collapse, it is of course necessary to
understand how to incorporate matter self-consistently into the
model. It is also important to establish whether
other
2-D models (such as the one used in Ref.\cite{CGHS}) can be
quantized
using these techniques. Finally,  one
would like to have a better understanding of the derived
probability
amplitude and its physical interpretation in the context of
quantum
gravity. These questions are currently under investigation.  \par
\vspace*{20pt} {\Large\bf Acknowledgements} G.K. is grateful to
M. Carrington, R. Epp,  P.Kelly, R. Kobes,
E. Martinez and D.
Vincent
for helpful discussions, and to K. Mak for producing the
diagrams.
This work was supported in part by the Natural Sciences and
Engineering Research Council of Canada.  \par \clearpage \begin{center} {\bf Figure
Caption}
\end{center} \par\noindent Figure 1: Plot of the (unnormalized)
probability amplitude $P[M/m]$ for black holes of mass $m/m_{pl}=
\{.1\,, .5\,, 2\,, 10\}$, showing the sharp decrease in width as
$m$
increases.  
\begin{thebibliography}{99}
\bibitem{dimensions}In order to make $\psi$ dimensionless one can
introduce an arbitrary dimensionful parameter $\lambda_0$ into
the
parametrization. However, this parameter has no physical
significance
and drops out of the final expressions.
\bibitem{quantum hair}It is
interesting
to note that for configurations in which the scalar field $\tau$
asymptotically approachs $\half r^2(1+\epsilon/r)$, the ADM mass
becomes:  $H_{ADM}= m - \epsilon/4$.
\bibitem{technical stuff}  An interesting feature of the present
model is the fact\cite{GK2} that
the choice of functional measure in \req{eq: measure}
has no
effect on the final
probability amplitude described by physical quantum states as
long as the momentum operators are formally self-adjoint with
respect to that measure.
\bibitem{mess}  It is shown in \cite{GK2} that the measure
induced from \req{eq: measure} on the space of $M$ is trivial, up
to formally infinite contributions that can be regulated to zero.
\end{thebibliography}
\end{document}